\documentclass[a4paper,reqno,12pt,final]{amsart}
\usepackage{amssymb,euscript,bbold}
\usepackage{array}
\usepackage[notref,notcite,color]{showkeys}
\usepackage{graphicx}
\usepackage{subfigure}
\renewcommand{\d}{\partial}
\newcommand{\half}{\tfrac12}
\renewcommand{\gg}{\mathfrak g}

\newcommand{\D}{\mathsf{D}}
\newcommand{\1}{\mathbb 1}

\newcommand{\Z}{\mathbb Z}

\newcommand{\eC}{\EuScript{C}}
\newcommand{\eB}{\EuScript{B}}

\newcommand{\Aut}{\mathrm{Aut}}
\newcommand{\Inn}{\mathrm{Inn}}
\newcommand{\Out}{\mathrm{Out}}

\newcommand{\U}{\mathrm{U}}
\newcommand{\SU}{\mathrm{SU}}

\DeclareMathOperator{\Ad}{Ad}

\begin{document}

\title[]{A note on D-branes in group manifolds: flux quantisation and
D0-charge}
\author[]{Sonia Stanciu}
\address[]{\begin{flushright}Spinoza Institute\\
Utrecht University\\
Leuvenlaan 4\\
3508 TD Utrecht\\
The Netherlands\\
\end{flushright}}
\email{s.stanciu@phys.uu.nl}
\thanks{${}^*$ Spin-00/17, {\tt hep-th/0006145}}
\begin{abstract}
  We show that a D-brane in a group manifold given by a (twisted)
  conjugacy class is characterised by a gauge invariant two-form field
  determined in terms of the matrix of gluing conditions.  Using a
  quantisation argument based on the path integral one obtains the
  known quantisation condition for the corresponding D-branes.  We
  find no evidence for the existence of a quantised $\U(1)$ gauge
  field flux.  We propose an expression for the D0 charge of such
  D-branes.
\end{abstract}
\maketitle

\section{Introduction}

Recently the issues of D0 charge and $\U(1)$ flux quantisation for a
class of D-branes in the $\SU(2)$ group manifold have attracted a
great deal of attention \cite{BDS,Pw,Ta,KKZ,AMM,Marolf1}.  Our aim
here is to show, using a somewhat different line of argument, that the
same basic results can be obtained without any reference to a
hypothetical $\U(1)$ gauge field or, indeed, of its flux.

In this letter we show, using the formalism developed in
\cite{SDnotes}, that a D-brane in a group manifold sitting on a
(twisted) conjugacy class $\eC$, and described, in the framework of
the boundary state approach, by the matrix $R$ of gluing conditions is
characterised by a gauge invariant two-form field $\omega$ defined on
the worldvolume of the D-brane whose components are determined by $R$.
By comparing the boundary conditions coming from the gluing conditions
with the ones deduced from the classical sigma model action, we are
able to identify this two-form field $\omega$ with the gauge-invariant
combination $B+2\pi\alpha'F$.

In order to write the boundary WZW action in terms of the three-form
$H$ and the two-form $\omega$ one is forced to introduce, much as in
the case of the standard WZW model, a certain field extension $\Tilde
g$.  The requirement that the quantum theory be independent of the
choice of field extension imposes two quantisation conditions
\cite{KlS,Gaw}: the first one, imposed on $H$ alone and similar to the
closed string case, is an integrality condition on the cohomology
class of $H$ in $H^*(\mathbf{G})$; whereas the second is that the
periods of $(H,\omega)$ over cycles in the relative homology
$H_3(\mathbf{G},\eC)$ take integer values.  In the case of $\SU(2)$ at
level $k$, these conditions yield $k+1$ D-branes: two point-like
D-branes situated at the two elements in the centre of $\SU(2)$, and
$k-1$ spherical D2-branes \cite{AS}.

A closer look at this quantisation condition suggests a natural
definition for the D0-brane charge of a given D-brane, which reads
\begin{equation}
  \label{Q0}
  \frac{1}{4\pi^2 T} Q_0 = \frac{1}{2\pi} \left(\int_{\d\eB} \Tilde
  g^*\omega  - \int_\eB \Tilde g^* H\right) \qquad \pmod{k}~,
\end{equation}
where $\eB$ is a three-manifold such that $g(\d \eB) = \eC$.  This
quantity is naturally gauge invariant and quantised (with integer
values), independently of any assumption regarding the existence of a
$\U(1)$ gauge field on the brane.  In the particular case of the
D2-branes in $\SU(2)$, the $H$ field contribution is similar to the
Poynting-type bulk contribution advocated in \cite{Ta}, and is valid
also when $H$ belongs to a nontrivial cohomology class.  This quantity
can be thought of as a generalisation of the $\U(1)$ flux in the case
where the three-form field $H$ belongs to a nontrivial cohomology
class.

\section{Semi-classical analysis}

\subsection{Boundary conditions from the boundary state approach}

We consider D-branes in a group $\mathbf{G}$ which preserve conformal
invariance and the infinite-dimensional symmetry of the current
algebra of the bulk theory.  They are described in terms of the
following gluing conditions:
\begin{equation}
  \label{eq:gc}
  J = R\bar J~,
\end{equation}
where the matrix of gluing conditions $R:\gg\to\gg$ is a
metric-preserving automorphism of the Lie algebra $\gg$; that is,
\begin{equation}
  \label{eq:Rhom}
  [R(T_a),R(T_b)] = R([T_a,T_b])~,
\end{equation}
and
\begin{equation}
  \label{eq:Rmet}
  R^T\Omega T = \Omega~,
\end{equation}
in the obvious notation.  This type of gluing conditions describe
\cite{AS,FFFS,SDnotes} D-branes whose worldvolumes lie on twisted
conjugacy classes.  More precisely, $\D$-branes in a WZW model with
group $\mathbf{G}$ come in several types, classified \cite{FSNW} by
the group $\Out_o(\mathbf{G})$ of metric-preserving outer
automorphisms of $\mathbf{G}$, which is defined as the quotient
$\Aut_o(\mathbf{G})/\Inn_o(\mathbf{G})$ of the group of
metric-preserving automorphisms by the invariant subgroup of inner
automorphisms.

In \cite{SDnotes} it was shown that the above gluing conditions give
rise, at a given point $g$ in $\mathbf{G}$, to the following boundary
conditions
\begin{equation*}
  \d g =  \mathbf{R}(g) \bar\d g~,
\end{equation*}
where the map $\mathbf{R}(g): T_g\mathbf{G} \to T_g\mathbf{G}$ is
defined as
\begin{equation*}
  \mathbf{R}(g) = - (\rho_g)_* \circ R \circ (\lambda_g)_*^{-1}~.
\end{equation*}
For the purposes of this paper it will be convenient to write the
above boundary conditions in a different form.  We therefore
parametrise $\mathbf{G}$ by introducing the coordinates $X^{\mu}$,
with $\mu=1,...,\dim\mathbf{G}$; we also introduce the left- and
right-invariant vielbeins
\begin{equation*}
  g^{-1}dg = {e^a}_{\mu}~dX^{\mu}T_a
  \qquad\text{and}\qquad
  dg g^{-1} = {\bar e}^a{}_{\mu}~dX^{\mu}T_a~.
\end{equation*}
These vielbeins are related by
${\bar e}^a{}_{\mu} = {e^b}_{\mu}{A^a}_b$, where $A$ denotes the
adjoint action of the group: $g T_a g^{-1} = {A^b}_a T_b$.  Using this
set-up, one can easily see that the gluing conditions \eqref{eq:gc}
give rise to the following boundary conditions for the component
fields $X^{\mu}$:
\begin{equation*}
  \d X^{\mu} = {\Tilde R(g)}^{\mu}{}_{\nu} \bar\d X^{\nu}~,
\end{equation*}
where the matrix of boundary conditions $\Tilde R(g)$ is given by
\begin{equation}\label{eq:Rtilde}
\Tilde R(g) = -{\bar e}^{-1} R e~.
\end{equation}
A Dirichlet direction is determined by an eigenvector of $\Tilde R(g)$
with eigenvalue $-1$, whereas all the other eigenvectors correspond to
Neumann directions, that is, directions tangent to the worldvolume of
the D-brane.

If we parametrise the worldsheet of the string by $(\sigma,\tau)$ we
can rewrite the above boundary conditions in the following form
\begin{equation}
  \label{eq:bc}
  i(\1 + \Tilde R)\d_{\sigma}X = (\1 - \Tilde R)\d_{\tau}X~,
\end{equation}
where $\d,\bar\d=\d_{\tau}\mp i\d_{\sigma}$.  We know that in this
case the worldvolume of a D-brane passing through $g$ and being
described by \eqref{eq:gc} is given by the twisted conjugacy class
$\eC_R(g)$.  We therefore consider the following split
\begin{equation}\label{eq:split}
  T_g\mathbf{G} = T_g \mathbf{G}^{||} \oplus T_g \mathbf{G}^{\perp}~,
\end{equation}
of the tangent space of $\mathbf{G}$ at the point $g$, where
$T_g \mathbf{G}^{||}$ is the tangent space to the twisted conjugacy
class, and $T_g \mathbf{G}^{\perp}$ is its orthogonal complement.
Using this, one can split the boundary conditions \eqref{eq:bc} into
two sets of conditions:
\begin{align*}
  i(\1 + \Tilde R)\d_{\sigma}X^{||} &= (\1 -
  \Tilde R)\d_{\tau}X^{||}~,\\ 
  i(\1 + \Tilde R)\d_{\sigma}X^{\perp} &= (\1 -
  \Tilde R)\d_{\tau}X^{\perp}~,
\end{align*}
in the obvious notation.  Since
$\left.\Tilde R\right|_{T_g \mathbf{G}^{\perp}} = -\1$,
from the second equation above we obtain the Dirichlet boundary
conditions
\begin{equation*}
\d_{\tau}X^{\perp} = 0~.
\end{equation*}
On the other hand, by using the fact that
$\left.(\1+\Tilde R)\right|_{T_g \mathbf{G}^{||}}$ is invertible, we
obtain the Neumann boundary conditions
\begin{equation*}
  \d_{\sigma}X^{||} + i\ \frac{\1 - \Tilde R(g)}{\1 +
    \Tilde R(g)}\ \d_{\tau}X^{||} = 0~,
\end{equation*}
We will now show that the matrix which defines the above Neumann
boundary conditions coincides with the one defining the two-form
$\omega$ on the worldvolume of the D-brane.

\subsection{Boundary conditions from the sigma model}

In the next section we will briefly review the definition of the
boundary WZW model.  In particular we will see that the action
\eqref{eq:owzw'} of a generic WZW model on a 2-space with a disc
topology is specified in terms of the three-form field $H$, familiar
from the standard case (when the worldsheet has no boundary), and a
two-form field $\omega$ defined on the worldvolume $\eC$ of the
D-brane, and satisfying $d\omega=\left.H\right|_\eC$.

The infinitesimal variation of the boundary WZW action contains a
bulk term (yielding the same equations of motion as in the closed
string case) and a boundary term which reads
\begin{equation*}
  \left. \int_{\d\Sigma} d\tau (g^{-1}\delta g)^a \left[ G_{ab}
      (g^{-1} \d_{\sigma} g)^b -i (g^*\omega)_{ab} (g^{-1}\d_{\tau}
      g)^b\right]\right|^{\sigma=\pi}_{\sigma=0}
\end{equation*}
where we have denoted by $G$ is the bi-invariant metric on the group
manifold.

Here we are interested in D-branes described by (twisted) conjugacy
classes.  Thus, in order to separate the Neumann and Dirichlet
boundary conditions encoded in the boundary term above, we must make
use of the specific form of a conjugacy class.  We recall (for details
see, e.g., \cite{SDnotes}) that this is defined as
\begin{equation*}
\eC_R(g_0) = \left\{ g=r(h)g_{0}h^{-1} \mid h\in\mathbf{G}\right\}~,
\end{equation*}
where the map $r:\mathbf{G} \to \mathbf{G}$ is defined by
\begin{equation*}
r\left(e^{tT}\right) = e^{tR(T)}~,
\end{equation*}
for $t$ small enough and $T$ any element in the Lie algebra.

Hence in this case $g$ maps the boundary of the worldsheet
$\d\Sigma$ into the conjugacy class $\eC_R(g_0)$ that is,
$g(\d\Sigma)\subset \eC_R(g_0)$, and therefore
\begin{equation*}
  \left.g^{-1}\delta g\right|_{\eC_R(g_0)} = (\Ad_{g^{-1}} R - \1)
  \delta hh^{-1}~. 
\end{equation*}
Assuming that the metric restricts nondegenerately to the worldvolume
of the D-brane (this is only a restriction in pseudo-riemannian
signature), then the infinitesimal variation $g^{-1}\delta g$ can be
written as the sum of two terms given by
\begin{align*}
  (g^{-1}\delta g)^{||} &=  (\Ad_{g^{-1}} R - \1)^{||} \delta
  hh^{-1}~,\\
  (g^{-1}\delta g)^{\perp} &= (\Ad_{g^{-1}} R - \1)^{\perp} \delta
  hh^{-1}~.
\end{align*}
Since $(\Ad_{g^{-1}} R - \1)^{\perp} = 0$, the second equation above
yields the Dirichlet boundary conditions
\begin{equation*}
  (g^{-1}\delta g)^{\perp} = 0~,
\end{equation*}
whereas the boundary term in the infinitesimal variation of the action
becomes
\begin{equation*}
  \int_{\d\Sigma} (\Ad_{g^{-1}} R - \1)^{||}(\delta hh^{-1}) \left[
    G(g^{-1}\d_{\sigma} g) -i (g^*\omega) (g^{-1}\d_{\tau}
    g)\right]^{||}~.
\end{equation*}
Taking into account that $(\Ad_{g^{-1}} R - \1)^{||}$ is
nondegenerate, we obtain the Neumann boundary conditions
\begin{equation*}
  (g^{-1}\d_{\sigma} g)^{||} -i G^{-1}(g^*\omega)(g^{-1}\d_{\tau}
  g)^{||} = 0~.
\end{equation*}
If we now consider the field $g$ to be parametrised as in the previous
paragraph, we can rewrite the Dirichlet and Neumann boundary
conditions in the following form
\begin{equation*}
  \delta X^{\perp} = 0~,
\end{equation*}
\begin{equation*}
  \d_{\sigma} X^{||} -i {\Tilde G}^{-1}\Tilde \omega \d_{\tau} X^{||}
  = 0~,
\end{equation*}
where $\Tilde G=e^T Ge$, $\Tilde\omega=e^T(g^*\omega) e$.

\subsection{The two-form field $\omega$}

By identifying now the Neumann boundary conditions obtained from the
boundary state approach with the Neumann conditions obtained from the
classical sigma model, we can deduce that the two-form $\omega$ is
uniquely determined by the matrix of gluing conditions $R$.  Indeed we
first deduce that
\begin{equation*}
\Tilde\omega = -\half~\langle dX\ , \frac{\1 - \Tilde R(g)}{\1 + 
                \Tilde R(g)}\ dX\rangle~,
\end{equation*}
from where we finally obtain that
\begin{equation}\label{eq:Omega}
g^*\omega = -\half~\langle g^{-1}dg\ , \frac{\1 + \Ad_{g^{-1}}R}{\1 -
    \Ad_{g^{-1}}R}\ g^{-1}dg\rangle~.
\end{equation}
Notice that this form is well defined on $\eC_R(g_0)$, as
$\1 + \Tilde R(g_0)$ is invertible on $T_{g_0} \mathbf{G}^{||}$.  One
can easily check that $g^*\omega$ is antisymmetric, hence it does
indeed define a differential two-form.  We know that the basic
property that this field must satisfy is
\begin{equation}\label{eq:omcond}
d(g^*\omega) = \left. g^*H\right|_{\eC_{R(g_0)}}~,
\end{equation}
where $H$ is the WZW three-form.  In order to verify that the two-form
field defined in \eqref{eq:Omega} does indeed satisfy this property,
we use the fact that the left-invariant Maurer--Cartan form evaluated
on $\eC_R(g_0)$ reads
\begin{equation*}
  \left.g^{-1}dg\right|_{\eC_R(g_0)} = (\Ad_{g^{-1}} R - \1) dhh^{-1}~.
\end{equation*}
This allows us to evaluate $H$ on the conjugacy class
\begin{equation*}
\left. g^*H\right|_{\eC_R(g_0)} = - d\ \langle\ dhh^{-1},
                           \Ad_{g^{-1}}R(dhh^{-1})\ \rangle~.
\end{equation*}
As expected, we obtain that the three-form field $g^*H$ is trivial in
de Rham cohomology when restricted to the (twisted) conjugacy class.
Furthermore, for $\omega$ itself we obtain
\begin{equation*}
  g^*\omega = -~ \langle\ dhh^{-1}, \Ad_{g^{-1}}R(dhh^{-1})\ \rangle~.
\end{equation*}
We thus conclude that a D-brane configuration which is given by
\eqref{eq:gc} and described geometrically by a (twisted) conjugacy
class $\eC_R$ in a group manifold $\mathbf{G}$ is endowed with a
two-form field $\omega$ which is uniquely determined in terms of the
matrix of gluing conditions $R$.  This implies, in particular, that if
one makes a certain gauge choice for the $B$ field in the bulk, then
the field $F$ on a given D-brane is uniquely determined in terms of
$\omega$ and the pull-back on the worldvolume of the D-brane of that
$B$ field.




\section{Quantum considerations}

We recall that the WZW model is defined by the action
\begin{equation*}
  I[g] = \int_{\Sigma}\langle g^{-1}\d g,g^{-1}\bar\d g\rangle +
  \int_{\eB} {\Tilde g}^* H~,
\end{equation*}
where $G_{ab}=\langle T_a,T_b\rangle$ defines a bi-invariant metric on
the group manifold, $\Sigma$ is Riemann surface without boundary, and
$\eB$ is a three-manifold with boundary $\d \eB=\Sigma$.  As is well
known, the WZ term in this case is a nonlocal term, defined in terms
of an extension $\Tilde g:\eB\to \mathbf{G}$ of the map $g$, such that
${\Tilde g}|_{\d \eB=\Sigma}=g$, and given by
\begin{equation}
  \label{eq:H}
  {\Tilde g}^* H = -\frac{1}{3}\langle{\Tilde g}^{-1}d{\Tilde g},
  d({\Tilde g}^{-1}d{\Tilde g})\rangle~.
\end{equation}
Thus the WZ term depends on the choice of extension $\Tilde g$, which
introduces in the action $I[g]$ an ambiguity proportional to the
periods of $H$ over the integer homology $H_3(\mathbf{G})$.  At the
classical level these discrete contributions are not relevant, as they
do not affect the equations of motion.  However at the quantum level,
the requirement that the path integral be independent of the choice of
extension $\Tilde g$ will in general fix the metric.  In the case of
$\mathbf{G}$ a compact simple Lie group, the metric can be fixed
uniquely by the requirement that
\begin{equation*}
  \frac{1}{2\pi} \int_Z H = k~,
\end{equation*}
where $Z$ is a $3$-cycle in $\mathbf{G}$ representing the generator of 
$H_3(\mathbf{G})\cong \Z$, and $k$ is a positive integer (the level).

The boundary WZW model was analysed in some detail in \cite{KlS,Gaw};
here we review a few aspects of particular interest for our
discussion.  The classical theory is usually defined by an action 
\begin{equation}
  \label{eq:owzw}  
  S[g] = \int_{\Sigma} \langle g^{-1}\d g, g^{-1}\bar\d
  g\rangle + \int_{\Sigma} g^* B + \int_{\d\Sigma} g^* A~.
\end{equation}
In this case the worldsheet $\Sigma$ is a two-dimensional manifold
with boundary $\d\Sigma$, and $B$ represents a particular choice for
the antisymmetric tensor field, such that $dB=H$.  A D-brane
configuration is characterised in this setting by a two-form $\omega$
defined on its worldvolume $\eC$, and satisfying
$d\omega=\left.dB\right|_\eC=\left.H\right|_\eC$.  Since
$\left.d(B-\omega)\right|_\eC=0$, one can define locally the one-form
potential $A$ such that $dA=B-\omega$.

One can write the boundary WZW action a manifestly gauge invariant
form, by using the three-form $H$ and the two-form $\omega$.  Let us
assume, for simplicity, that we have one D-brane sitting on a
(twisted) conjugacy class $\eC$ in $\mathbf{G}$.  In this case the
worldsheet $\Sigma$ can be represented in terms of a closed surface
$\Sigma'$, where $\Sigma=\Sigma'\backslash D$, and $D$ is a (unit)
disk embedded in $\Sigma'$.  Provided that some topological
obstructions can be overcome (which amount to the vanishing of the
relative homology group $H_2(\mathbf{G},\eC)$), one can then extend
$g$ to a map $g':\Sigma'\to\mathbf{G}$ such that $g'(D)\subset \eC$,
and $g'$ can be further extended to a map
${\Tilde g}':B'\to\mathbf{G}$, with $B'$ a three-dimensional manifold
such that $\d B'=\Sigma'$.  This allows us to write the WZ term in a
more familiar form
\begin{equation}
  \label{eq:owzw'}  
  S[g] = \int_{\Sigma} \langle g^{-1}\d g, g^{-1}\bar\d
  g\rangle + \int_{B'} {\Tilde g}^{'*} H - \int_{D} g^{'*}\omega~.
\end{equation}
Thus, in this case, the WZ term has a bulk component and a boundary
component (defined on the worldvolume of the D-brane).  Similarly to
the standard case, the WZ term depends on the extension $\Tilde g'$,
which introduces an ambiguity in the action
\begin{equation}
  \label{eq:amb}
  \left(\int_{\Tilde \eB} {\Tilde g}^* H - \int_{S^2} {\Tilde
      g}^*\omega\right)~,
\end{equation}
where $\Tilde \eB$ is a three-dimensional manifold with
$\d\Tilde \eB=S^2$ and $\Tilde g:\Tilde \eB\to\mathbf{G}$ such that
$\Tilde g(S^2)\subset \eC$.  As shown in \cite{KlS,Gaw} these are
proportional to the periods of $(H,\omega)$ over the cycles of the
relative homology $H_3(\mathbf{G},\eC)$.

In order to evaluate this ambiguity and compare it to the standard
case without boundary it is convenient to ``fill'' $\Tilde \eB$ with
the unit ball $\eB$ (whose boundary is $S^2$), ending up with a
three-dimensional manifold $\Hat \eB$ without boundary; if we also
extend $\Tilde g$ to a map $\Hat g:\Hat \eB\to\mathbf{G}$, we can
rewrite \eqref{eq:amb} as the sum of two terms, where the first one
\begin{equation}
  \label{eq:amb1}
  \int_{\Hat \eB} {\Hat g}^* H~,
\end{equation}
has the same form as the ambiguity appearing in the standard WZW
action.  This fixes the metric just as in the previous case, to ensure
that \eqref{eq:amb1} induces no dependence on our field extensions at
the level of the path integral.  This leaves us with the second term
\begin{equation}
  \label{eq:amb2}
  \left(\int_{\d \eB} {\Tilde g}^*\omega - \int_{\eB}{\Tilde g}^*
  H\right)~.
\end{equation}
which is characteristic to the boundary WZW model.  Hence if we want
that the path integral be independent of $\Tilde g$, this term must
take values in $2\pi\Z$.  This can be thought of as a generalisation
of the Dirac quantisation condition.

\section{The $\SU(2)$ case}

\subsection{(Semi-)classical analysis}

Let us now apply the above discussion to the particular case of
D-brane configurations given by conjugacy classes in $\SU(2)$.  We use
the following parametrisation \cite{BDS}:
\begin{equation*}
  g=e^{i(\psi_1\sigma_1+\psi_2\sigma_2+\psi_3\sigma_3)}~,
\end{equation*}
where $(\psi_1,\psi_2,\psi_3)$ forms a vector of length $\psi$
pointing in the direction $(\theta,\phi)$, and
$(\sigma_1,\sigma_2,\sigma_3)$ are the Pauli matrices.  This
parametrisation, whose spacetime fields are $(\psi,\theta,\phi)$, has
the advantage that one of the coordinates, namely $\psi$, corresponds
to the Dirichlet direction, as we will see explicitly in a moment.  We
can compute, as usual, the invariant vielbeins $e$ and $\bar e$,
the sigma model metric $G$, and the Wess--Zumino three-form $H$ thus
obtaining
\begin{equation*}
  G = \frac{k}{2\pi}\left( d\psi^2 + \sin^2\psi (d\theta^2 +
  \sin^2\theta d\phi^2)\right)~,
\end{equation*}
\begin{equation*}
  H = \frac{k}{\pi} \sin\theta\, \sin^2\psi\, d\psi \wedge d\theta
  \wedge d\phi~,
\end{equation*}
where the level $k$ is a positive integer.  Furthermore, by using
\eqref{eq:Rtilde}, we obtain the matrix of boundary conditions
\begin{equation*}
  \Tilde R(g) =
  \begin{pmatrix}
    -1 & 0 & 0\\
    0 & - \cos 2\psi & - \sin\theta \sin 2\psi\\
    0 & \sin 2\psi \csc\theta & -\cos 2\psi
  \end{pmatrix}.
\end{equation*}
From the form of this matrix we can immediately read off that we
always have a Dirichlet boundary condition along the ``radial''
coordinate $\psi$.  In other words, the D-branes described by these
gluing conditions are normal to the $\psi$ direction at any given
point.  On the other hand, we know that these D-branes are conjugacy
classes in $\SU(2)$---hence every such conjugacy class $\eC=\eC(\psi)$
is a two-sphere centred around the identity.  In particular, for
$\psi=0,\pi$ we get the zero-dimensional D-branes since $\Tilde
R=-\1$.

According to the discussion in Section 2, we can now calculate
the two-form field $\omega$ associated to a given D-brane $\eC(\psi)$
obtaining
\begin{equation*}
  \Tilde\omega = -\frac{k}{4\pi} \sin 2\psi\, \sin\theta \, d\theta
  \wedge d\phi~.
\end{equation*}
Using the explicit knowledge of this field, we can evaluate the energy
of such a configuration from the Born--Infeld action 
\begin{align}
  \label{eq:energy}
  E(\psi) &= 2\pi T \int_{\eC(\psi)}\sqrt{\det(\Tilde g
             +\Tilde\omega)}\nonumber\\
          &= 4\pi k T \sin\psi~,
\end{align}
where we denoted by $T$ the D-brane tension, and by $\Tilde g$ the
metric induced on the worldvolume of the D-brane.  From this
expression we can immediately see that the energy of such a D-brane
configuration reaches its minimum for $\psi=0,\pi$, i.e., for the
zero-dimensional D-branes.  Hence, from a classical point of view, it
is only the two D0-branes that give rise to stable configurations.

Let us compare our result with the one obtained in \cite{BDS}.  The
main difference between the two approaches lies in the way one
determines the $B+2\pi\alpha'F$ field.  In \cite{BDS} some gauge
choices were involved.  Here, this field was determined uniquely, by
identifying the boundary conditions coming from the gluing conditions
with the ones of the classical sigma model, and thus the expression
for the energy \eqref{eq:energy} holds independently of any particular
gauge choice.  One could argue that the only necessary conditions are
that $dB=H$ and $dF=0$.  However, as we showed in Section 2,
consistency between the gluing conditions and the classical sigma
model boundary conditions constrains $B+2\pi\alpha'F$ to be equal to
the two-form field $\omega$.  One might also expect that the energy
minimisation procedure itself selects the right combination, but the
different results obtained in the two approaches indicate that this is
not the case.  It is perhaps useful to remark that  the
gluing conditions fix the shape of the D-brane (spherical in this
case) whereas minimising the energy basically fixes its size.
Moreover, there is an infinite number of D2-branes which, despite the
fact that they satisfy the gluing conditions, do not minimise the
Born--Infeld action.

\subsection{Quantum analysis}

Let us now apply the quantum considerations of the previous section to
our D-branes $\eC(\psi)$.  To this end, let us compute the period of
$(H,\omega)$ over a cycle in $H_3(\mathbf{G},\eC)$.  In this case
$\eB=\eB(\psi)$ is a three-ball bounded by $\eC(\psi)$ and we
calculate
\begin{equation*}
  \left(\int_{C(\psi)}{\Tilde g}^*\omega - \int_{\eB(\psi)}{\Tilde
  g}^* H\right) = -2k\psi~.
\end{equation*}
Hence, in order for the path integral to be independent of our field
extensions, $\psi$ must be quantised as follows
\begin{equation*}
\psi_n = \frac{n\pi}{k}~,\qquad\qquad n=0,1,...,k~.
\end{equation*}
This result, which agrees with the analysis of \cite{AS} (for a
detailed exposition see also \cite{Gaw}), allows us to conclude that
the $k+1$ D-branes singled out in this fashion are stable.  Moreover
if we evaluate their masses by using the Born--Infeld action, we
obtain
\begin{equation*}
  M_n = 4\pi k T\sin\left(\frac{n\pi}{k}\right)~,\qquad\qquad
  n=0,1,...,k~,
\end{equation*}
which, as pointed out in \cite{BDS}, agrees with the CFT calculations.

Such a quantisation condition appears to be a non-local condition
imposed on $\psi$, and concerns about its physical meaningfulness are
well founded\footnote{I thank M~Douglas for raising this point.}.
It seems clear that the reason why $\psi$ is determined to a
particular value is due to the fact that we are analysing a very
symmetric type of D-brane configurations, namely those given by
conjugacy classes.  Presumably, it is possible to deform the D-brane
while preserving the charge and the mass and while respecting
conformal invariance of the boundary conditions in such a way that the 
condition on $\psi$ gets smeared.

\subsection{D0 charge}

These results prompt us to propose the following definition for the
D0-charge of such D-branes:
\begin{equation}\label{eq:Q0}
  \frac{1}{4\pi^2 T} Q_0 = \frac{1}{2\pi} \left(\int_{\d \eB}{\Tilde
  g}^*\omega - \int_{\eB}{\Tilde g}^* H\right) \qquad \pmod{k}~.
\end{equation}
In the particular case of a D-brane given by $\eC(\psi)$ one obtains
\begin{equation*}
Q_0(\psi) = - 4\pi k T\psi\qquad \pmod{k}~,
\end{equation*}
which takes integer values for the $k+1$ D-branes obtained before.

We remark here that these values of the D0 charge appear to be
different from the results one obtains from the CFT calculations.  On 
the other hand, if we compute the flux of $\omega$ alone, one obtains 
\begin{equation}\label{eq:Q}
  Q^{\text{BDS}}(\psi) =  2\pi T\int_{\eC(\psi)}{\Tilde g}^*\omega = -
  2\pi k T\sin 2\psi~,
\end{equation}
which agrees, in the case of the $k+1$ stable configurations, with
\cite{BDS} and with the CFT results.  This seems to indicate that the
path integral and the boundary state approach compute two distinct
quantities, as suggested recently in \cite{Marolf1} (we will come back
to this point).  Notice that in the regime where $n\ll k$ one obtains
$Q_0(n)\simeq Q^{\text{BDS}}(n) \simeq M_n$, which is nothing but the
mass and charge of $n$ D-particles.

This definition for the D0-brane charge has the following virtues: it
is manifestly gauge invariant, as the one introduced in \cite{BDS},
but unlike the one based on the flux of the $\U(1)$ field
\cite{Ta,AMM}.  It is naturally quantised with integer values, as is
natural to expect of a RR charge.  Moreover, it includes a
contribution coming from the bulk field $H$, similar to the one
advocated in \cite{Ta}.  Notice however that this bulk term does not
cancel the $B$ field contribution included in the flux of $\omega$.
It is perhaps useful to discuss this point also in the framework used
in \cite{Ta}, where the Poynting-type contribution to the D0-brane
charge reads
\begin{equation*}
\frac{1}{6}\int G^{(4)}_{0ijk} H^{ijk}~,
\end{equation*}
with obvious notation.  In evaluating this contribution we must take
into account that, in this case, $H$ belongs to a nontrivial
cohomology class and hence there is no globally defined gauge
invariant $B$ field\footnote{A similar observation was made
  independently by A Tseytlin.}.  Therefore this results in a bulk
contribution which agrees with the bulk term in our definition of
$Q_0$ in \eqref{eq:Q0}.

Last but not least, notice that, although the relative cohomology
class of $(H,\omega)$, on which the definition of $Q_0$ is based, is
not the same as the flux of the gauge field $F$ defined on the
conjugacy class, it can nevertheless be thought of as a natural
generalisation of the $\U(1)$ flux in the case of a D-brane in a WZW
background, since it reduces to this in the particular case where $H$
is an exact form, as one can verify by using the Stokes theorem in the
second term in \eqref{eq:Q0}.

\section{Discussion}

In this letter we have analysed a class of D-brane configurations in
group manifolds, which is characterised by gluing conditions that
preserve the maximum possible amount of symmetry of the bulk theory,
namely, the current algebra of the WZW theory.  We know that the
gluing conditions generally fix the ``shape'' of a D-brane; in
particular, this type of gluing conditions give rise to D-branes
described by (twisted) conjugacy classes.  Here we have shown that
consistency between the gluing conditions and the sigma model boundary
conditions also fixes the gauge invariant field $B+2\pi\alpha'F$.
Using this fact, one can estimate the energy of such a D-brane
configuration, from the Born--Infeld action, independently of any
particular gauge choice for either $B$ or $F$.  One thus obtains that
in the $\SU(2)$ case, at the classical level, it is only the two
D0-branes that are stable.  We have then used a quantisation argument
based on the path integral which requires that the periods of
$(H,\omega)$ over the cycles of the relative homology
$H_3(\mathbf{G},\eC)$ take integer values; in the $\SU(2)$ case this
produces a discrete set of allowed D-brane configurations, whose mass
spectrum agrees with the CFT calculations.

This quantisation argument also prompted us to make an alternative
proposal for the D0-brane charge of such D-branes, which differs from
the one introduced in \cite{BDS} by a bulk term, similar, yet not
identical, to the one advocated in \cite{Ta}.  We believe this to be a
natural definition for a number of reasons.  First of all it is
manifestly gauge invariant, and we consider this to be an important
feature, as both the boundary WZW model and the Born--Infeld action of
a D-brane in such a background are manifestly gauge invariant.  This
D0 charge is also naturally quantised, taking integer values, and this
is clearly a desirable feature for a RR charge.  Finally, this
definition constitutes a natural generalisation of the $\U(1)$ flux.

It is important to remark that this definition also raises a number of
interesting questions.  One of them is the apparent discrepancy
between $Q_0$ and the CFT results for the D0 charge.  More precisely,
it seems that what the boundary state calculates is the flux of
$\omega$ through the D-brane, without the $H$ bulk contribution.
According to a recent analysis \cite{Marolf1}, both these quantities
can be given a natural physical interpretation: the flux \eqref{eq:Q}
of the two-form field $\omega$ (gauge invariant but not quantised) can
be understood as the brane source charge, whereas $Q_0$ can be thought
of as the Page charge\footnote{I thank D~Marolf for this
  observation.}, which is gauge invariant and quantised.  Using this
analogy one can understand better the relation between these two
charges.  Indeed, the brane source charge is not conserved in general
and the non-conservation rule is encoded, in our case, in the relation
between $\omega$ and $H$ on the D-brane which is given by
\eqref{eq:omcond}.  Moreover, this very relation also suggests us the
way to modify the brane source charge in order to obtain a conserved
quantity, which is nothing but our $Q_0$.  This procedure is
reminiscent of the way one constructs the Page charge in supergravity
(see, e.g., \cite{Kelly}). Is is quite remarkable that the D0 charge
obtained in this way turns out to agree with the one obtained by
imposing the well-definedness of the boundary WZW path integral, and
it is this agreement that renders it quantised.

One problematic aspect of this approach is the fact that the
quantisation argument seems to impose a non-local condition on the
spacetime field $\psi$.  Notice however that the evaluation of the D0
charge was made for a specific class of D-branes, described by a very
special type of gluing conditions; it is possible that this non-local
condition will get smeared once we allow for more general D-brane
configurations.

Finally, one of the most important and intriguing conclusions of this
analysis is that there appears to be no evidence for the existence of
a quantised $\U(1)$ gauge field flux on this particular class of
D-branes.  By this we do not mean that we have a $\U(1)$ gauge field
whose flux is not quantised.  Rather, we have shown that one can
define and analyse the boundary WZW model, both at the classical and
at the quantum level, without ever having to introduce a $\U(1)$ gauge
field on the brane.  Using this approach we obtained a mass spectrum
for the allowed D-brane configurations which agrees with the one
obtained in the boundary state formalism, we recovered the D0 charge
of \cite{BDS} and its spectrum, and we were able to define a new D0
charge which is manifestly gauge invariant and quantised and modular,
with the periodicity given by the level of the WZW model.  These
results suggest that the role of the $\U(1)$ gauge field $F$ in the
case of a background with $B=0$ is taken by over by the globally
defined gauge invariant two-form field $\omega$ in the case of a
background characterised by a nontrivial $B$ field.

\section*{Acknowledgements}
I would like to thank C~Bachas, M~Douglas, K~Gaw\c{e}dzki, M~Kreuzer,
D~Marolf and C~Schweigert for useful remarks and criticism,
A~Recknagel, V~Schomerus and S~Theisen for their hospitality at
AEI~Potsdam and ESI~Vienna, where part of this work was done, and
especially JM~Figueroa-O'Farrill and A~Tseytlin for many insightful
discussions and correspondence, and for helping me fix a stubborn
factor of $2$ or two.

\providecommand{\href}[2]{#2}\begingroup\raggedright\endgroup

\end{document}